\documentclass{PoS}

\pdfoutput=1

\usepackage{amsmath}
\usepackage{amssymb}
\usepackage{fontenc}
\usepackage{times}
\usepackage{mathptmx}
\usepackage{graphicx}
\usepackage{accents}

\newcommand{\bc}{\begin{center}}
\newcommand{\ec}{\end{center}}
\newcommand{\be}{\begin{equation}}
\newcommand{\ee}{\end{equation}}
\newcommand{\bea}{\begin{eqnarray}}
\newcommand{\eea}{\end{eqnarray}}
\newcommand{\bi}{\begin{itemize}}
\newcommand{\ei}{\end{itemize}}
\newcommand{\bt}{\begin{tabular}}
\newcommand{\et}{\end{tabular}}
\newcommand{\Dlr}{\accentset{\smash[t]\leftrightarrow}{D}}

\def\ks{\kappa^{\rm S}}
\def\ksc{\kappa^{\rm S}_c}
\def\mps{m_{\rm PS}}
\def\mpi{m_{\pi}}
\def\mN{m_{\rm N}}
\def\mdelta{m_{\rm \Delta}}
\def\msbar{\overline{\rm MS}}

\title{Probing the chiral limit with clover fermions II:\\The baryon sector}

\ShortTitle{Probing the chiral limit with clover fermions II: The baryon sector}

\author{
  QCDSF/UKQCD Collaboration:
  Meinulf G\"{o}ckeler$^a$,
  Philipp H\"{a}gler$^b$,
  Roger Horsley$^c$,
  Yoshifumi Nakamura$^d$,
  Munehisa Ohtani$^a$,
  \speaker{Dirk Pleiter}$^{\,d}$,
  Paul~E.L.~Rakow$^e$,
  Andreas Sch\"{a}fer$^a$,
  Gerrit Schierholz$^{df}$,
  Wolfram Schroers$^d$,
  Hinnerk St\"uben$^g$ and
  James M.~Zanotti$^c$\\
  \llap{$^a$} Institut f\"ur Theoretische Physik, Universit\"at Regensburg,
              93040 Regensburg, Germany\\
  \llap{$^b$} Institut f\"ur Theoretische Physik T39, Physik-Department der
              TU M\"unchen, 85747~Garching, Germany\\
  \llap{$^c$} School of Physics, University of Edinburgh, Edinburgh EH9 3JZ,
              UK\\
  \llap{$^d$} John von Neumann-Institut f\"ur Computing NIC, Deutsches
              Elektronen-Synchrotron DESY, 15738 Zeuthen, Germany\\
  \llap{$^e$} Theoretical Physics Division, Department of Mathematical Sciences,
              University of Liverpool, Liverpool L69 3BX, UK\\
  \llap{$^f$} Deutsches Elektronen-Synchrotron DESY, 22603 Hamburg, Germany\\
  \llap{$^g$} Konrad-Zuse-Zentrum f\"ur Informationstechnik Berlin,
              14195 Berlin, Germany\\
  Email: \email{dirk.pleiter@desy.de}
}

\abstract{%
Algorithmic progress in recent years made it possible to simulate QCD
with $N_f=2$ flavours of $O(a)$-improved Wilson fermions at very light
quark masses. We present the current results for baryon spectrum states,
the nucleon axial coupling and the lowest moment of unpolarised nucleon
structure functions. Special emphasis is given to a comparison of our
calculations with results from chiral effective theories.

\vspace*{20mm}
\hspace*{-8mm}\texttt{Edinburgh 2007/29}\\
\hspace*{-8mm}\texttt{DESY 07/178}
}

\FullConference{The XXV International Symposium on Lattice Field Theory\\
		 July 30 - August 4 2007\\
		 Regensburg, Germany}

\begin{document}

\section{Introduction}

Thanks to recent algorithmic improvements and the availability of a new
generation of capability computers, simulations with dynamical Wilson
fermions can now be extended to much lighter quark masses and bigger lattices.
In recent years the QCDSF collaboration has significantly extended the
number of ensembles of gauge configurations with $N_f=2$ flavours of
$O(a)$-improved Wilson fermions.  These datasets cover a large range of
quark masses as well as different lattice spacings and volumes.
In principle, this puts us in the position to check for systematic
errors that affect essentially all lattice calculations, i.e.~finite
size corrections, discretisation errors and errors due to extrapolations
to the physical quark masses or the chiral limit.

For the analysis of both, finite size effects and the quark mass
dependence, results from chiral effective theories can be used
to correct the simulation results or to guide extrapolations.
Although the strategy of combining the results obtained from
lattice simulations and chiral effective theories has led to a
consistent picture for various observables, e.g.~the pion decay
constant \cite{gerrit}, the mass of the
nucleon \cite{qcdsf-mn}, the axial coupling of the nucleon
\cite{qcdsf-ga} or the nucleon electromagnetic form factors
\cite{qcdsf-ff}, this approach suffers from the paucity of lattice
results in the region where chiral perturbation theory is expected
to be applicable.

\begin{figure}[b]
\includegraphics[scale=0.355]{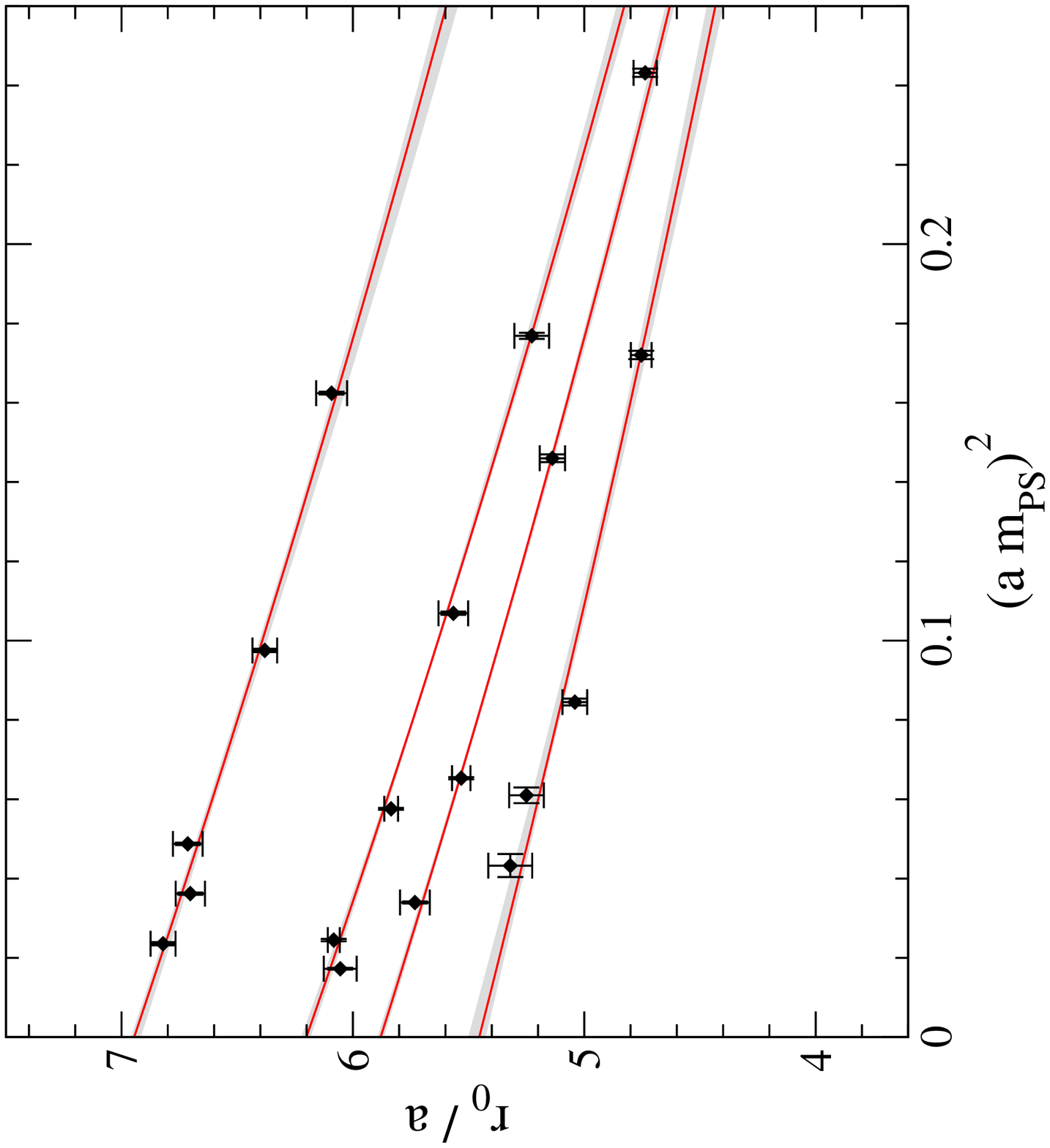}\hfill
\includegraphics[scale=0.36]{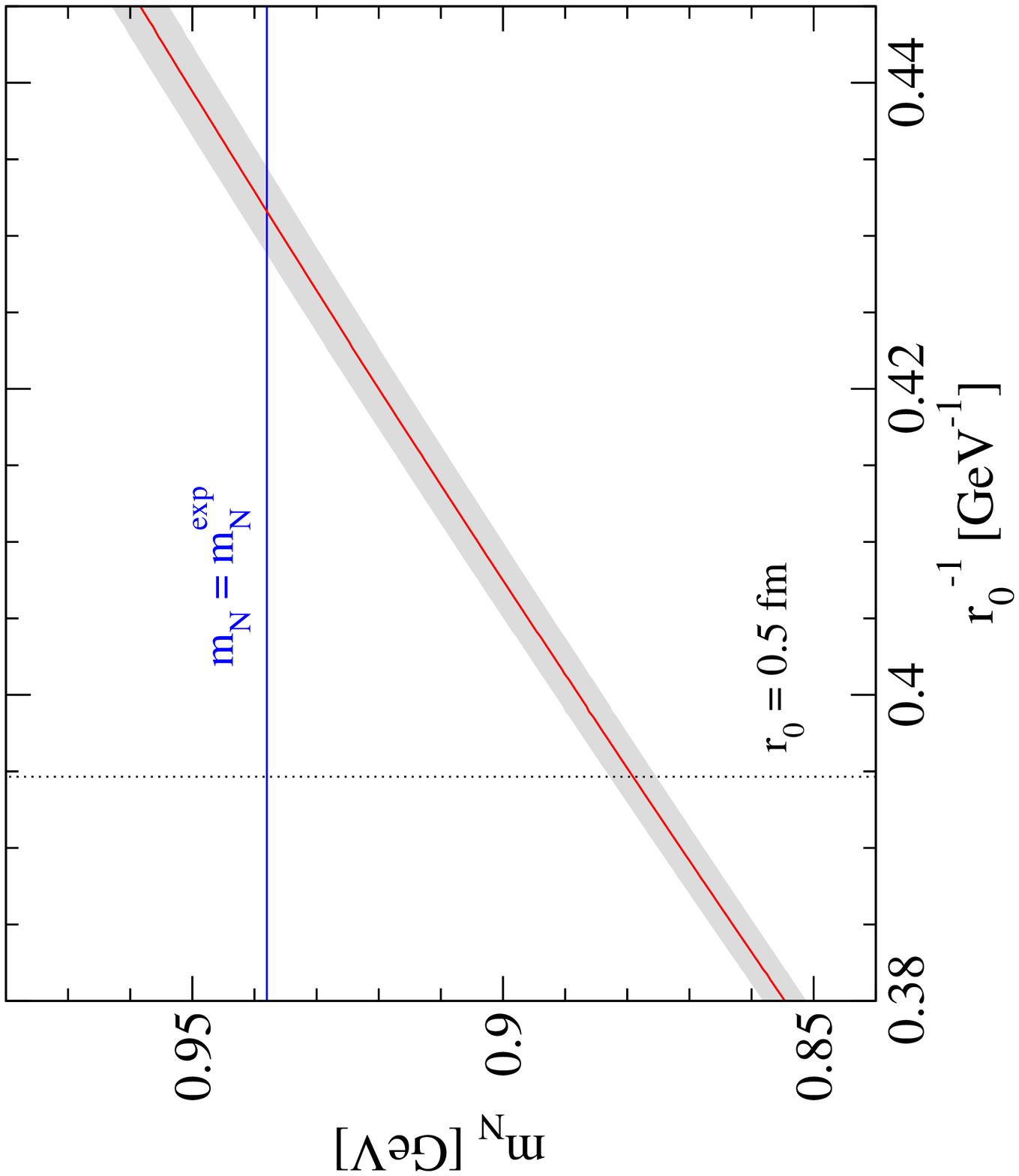}
\caption{\label{fig:r0}%
The left plot shows the values for $r_0(\beta,\ks)$ together with
a fit to Eq.~(\protect\ref{eq:r0}). In the right plot
$\mN(\mps=m_{\pi})$, which is determined from fits to
Eq.~(\protect\ref{eq:mn}) for different values of $r_0$,
is shown as a function of $r_0^{-1}$.}
\end{figure}

Another source for systematic errors stems from the translation of
the lattice results into physical units. A quantity which is
often used for this purpose is the Sommer scale $r_0$. While this quantity
has the advantage that on the lattice it can be determined with relatively
high statistical accuracy, its experimental value is far less well known.
To avoid depending on this experimental value, which relies on model
assumptions, one can alternatively request lattice results for a quantity
$X$, which has the dimension of a mass (e.g.~the pion decay constant or the
mass of the nucleon), to be equal to the corresponding experimental result,
i.e.~$r_0^{\rm lat}\,X^{\rm lat} = r_0\,X^{\rm exp}$.
The results from different collaborations suggest that the resulting
value for $r_0$ is significantly smaller than the typically
quoted experimental number $r_0 \simeq 0.5\,\mbox{fm}$.
In this paper we will use $r_0 = 0.467\,\mbox{fm}$
which allows for easy comparison with our previous results. Our
most recent results
for the nucleon mass (see next section and Fig.~\ref{fig:r0}) and
the pion decay constant (see \cite{gerrit}) suggest
that the actual value is even smaller.
To define our scale independent of the quark mass we extrapolate the
measured values of $r_0^{\rm lat}(\beta,\ks)$ to the chiral limit using
the ansatz
\be
\ln \frac{r_0^{\rm lat, c}}{a} =
A_{0}(\beta) + A_{2}(\beta) \left(a \mps(\beta,\ks)\right)^2\,,
\label{eq:r0}
\ee
where
$A_{i}(\beta) = A_{i0} + A_{i1} (\beta-\beta_0) + A_{i2} (\beta-\beta_0)^2$
and $\beta_0 = 5.29$.
We fit our data in the range $a\mps < 0.5$ and find
$\chi^2/N_{\rm d.o.f.} = 6.7/12$ (see Fig.~\ref{fig:r0}).

\section{Masses}

\begin{figure}[t]
\includegraphics[scale=0.55]{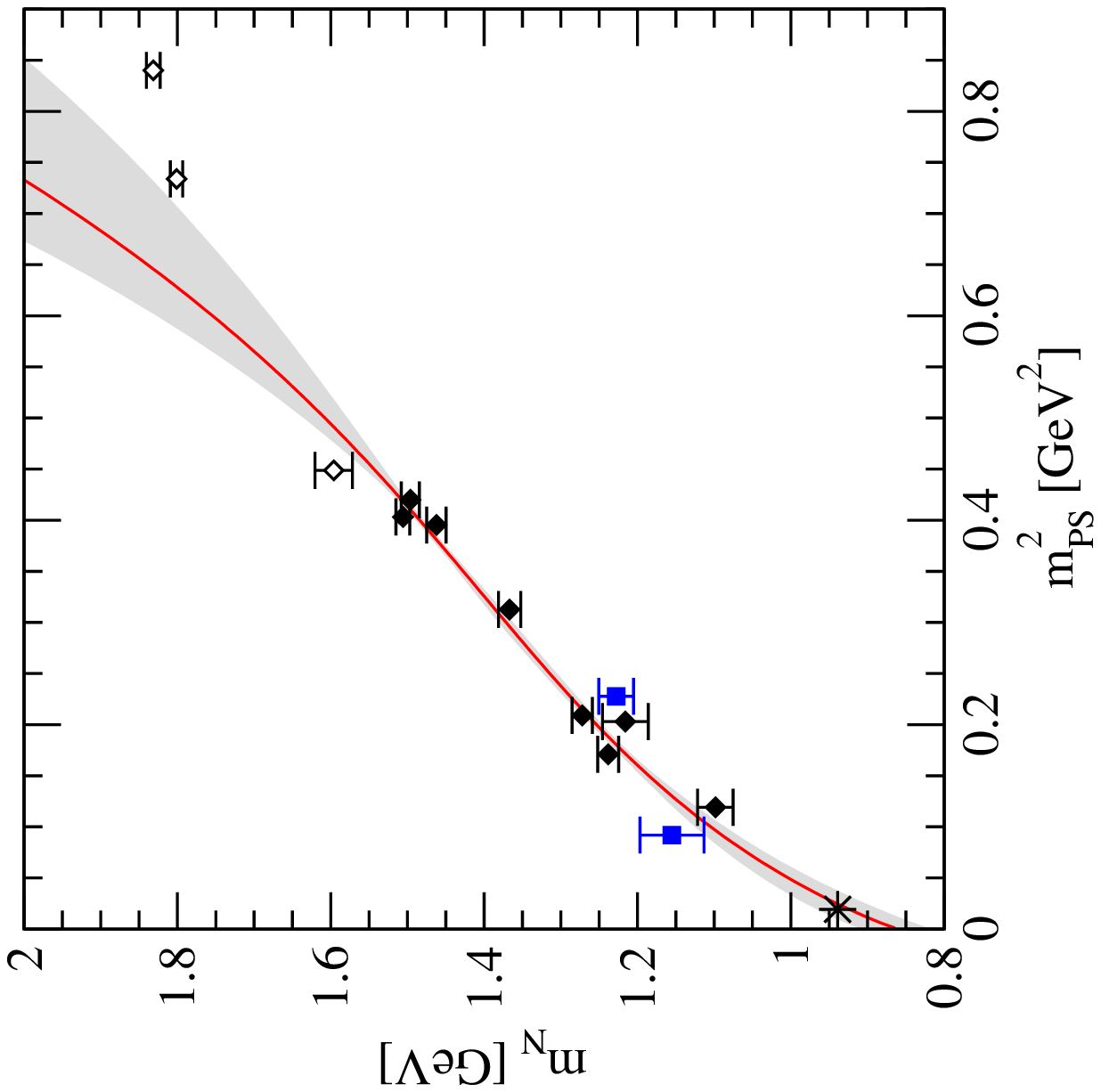}\hfill
\includegraphics[scale=0.55]{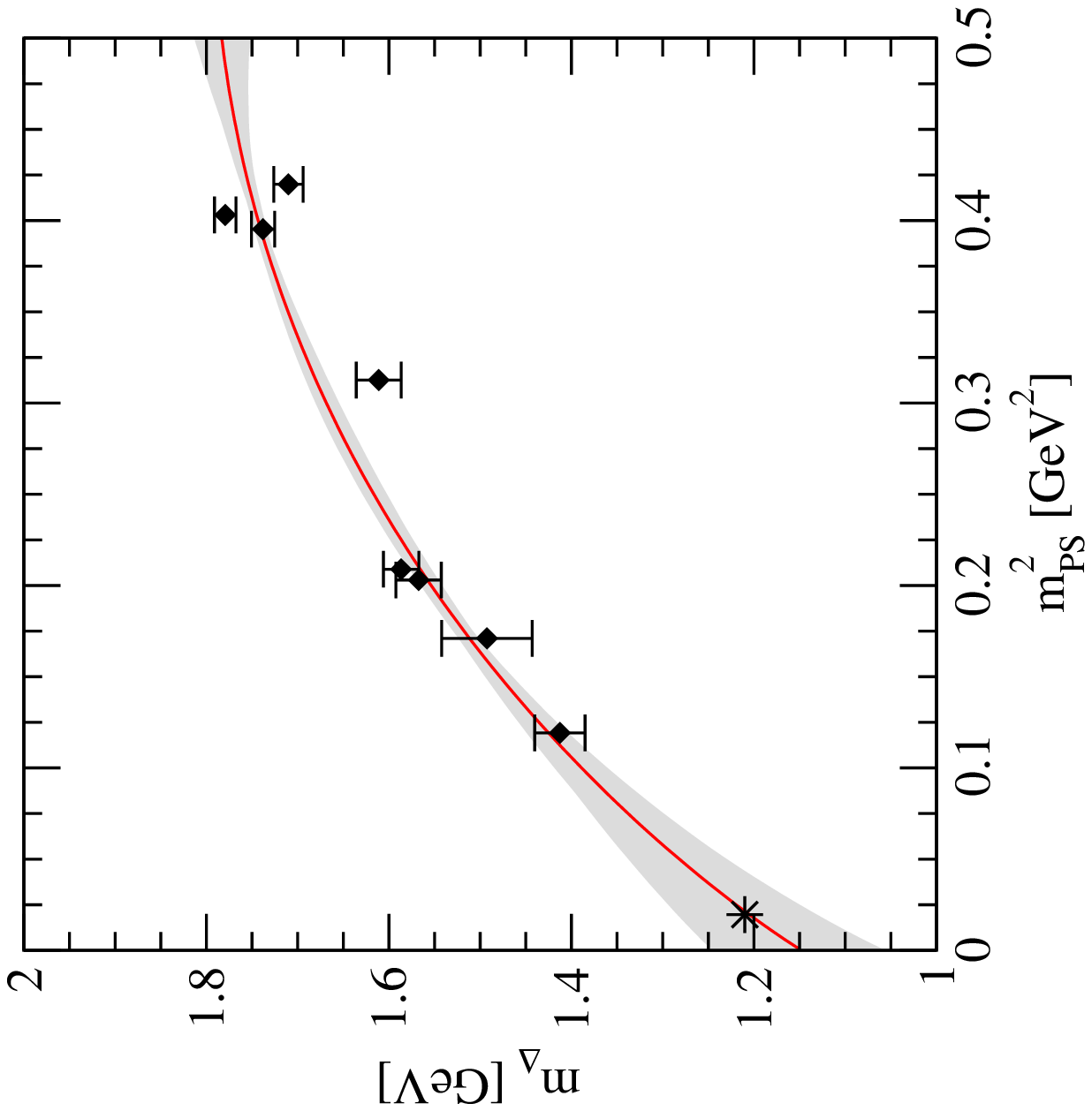}
\caption{\label{fig:mn_mdelta}%
Lattice results for $\mN$ (left) and $\mdelta$ (right) together with
a fit to Eq.~(\protect\ref{eq:mn}) and
Eq.~(\protect\ref{eq:mdelta}). The open symbols
indicate data which has not been included in the fit.
The filled squares symbols in the left plot show preliminary data on larger
lattices which has not been included in this analysis.}
\end{figure}

The mass of the nucleon, $\mN$, and delta, $\mdelta$, are
experimentally well-determined quantities. However, for both quantities
calculations using baryon chiral perturbation theory (B$\chi$PT) suggest 
a rather non-trivial quark mass dependence. In an infinite volume
$\mN$ depends on the quark mass in the following way \cite{qcdsf-mn}:
\begin{eqnarray}
\label{eq:mn}
 \mN(\mps) & = & {M_0} - 4 {c_1} \mps^2-
\frac{3 {g_{A,0}}^2}{32 \pi {F_0^2}} \mps^3 + \\\nonumber
&& \left[{e_1^r(\lambda)}-\frac{3}{64 \pi^2 {F_0}^2}
     \left( \frac{{g_{A,0}}^2}{M_0} -
     \frac{c_2}{2} \right) - \right.
\left.
   \frac{3}{32 \pi^2 {F_0}^2}
       \left( \frac{{g_{A,0}}^2}{M_0} - 8{c_1} +
       {c_2} + 4 {c_3} \right)
   \ln{\frac{\mps}{\lambda}} \right] \mps^4 \\\nonumber
&& + \;\;\frac{3 {g_{A,0}}^2}{256 \pi {F_0}^2 {M_0}^2}\mps^5 + O(\mps^6)\,.
\end{eqnarray}
Even extending our fit interval to $0 < \mps \lesssim 650\,\mbox{MeV}$
does not allow us to determine all parameters. We therefore restrict the
set of free fit parameters to the nucleon mass in the chiral limit
$M_0$, the not very well known low-energy constant (LEC) $c_1$ and the counter-term
$e_1^r(\lambda)$ (we use $\lambda = 1\,\mbox{GeV}$).
For the other parameters,
i.e.~the LECs $c_2$ and $c_3$, the pion decay constant $F_0$ and
the nucleon axial coupling $g_{A,0}$,
we use the phenomenological estimates 
listed in Table~\ref{tab:phen}. Our nucleon mass data and the fit
are shown in Fig.~\ref{fig:mn_mdelta}. Note that all our results seem to
fall on a universal curve, indicating discretisation effects to
be small. We therefore ignored $O(a^2)$ effects in our fit ansatz.
We observe that $\mN(\mps=m_{\pi})$ is consistent with experiment.
Furthermore, we find $c_1 = -1.02(7)\,\mbox{GeV}^{-1}$, a value
which is consistent with other estimates~\cite{Bernard:2007zu}.
We have repeated our analysis for different values for $r_0$.
The results are shown in Fig.~\ref{fig:r0}. If we set
$\mN = \mN^{\rm exp}$ we obtain $r_0 = 0.457(3)\,\mbox{fm}$.

\begin{table}[b]
\bc
\bt{|l|l|l||l|l|l||l|l|l|}
\hline
$c_2$       & $3.2\,\mbox{GeV}^{-1}$   & See \cite{qcdsf-mn} &
$\Delta_0$  & $0.271\,\mbox{GeV}$      & \cite{Eidelman:2004wy} &
$g_{A,0}$   & $1.2(1)$                 & \cite{Hemmert:2003cb} \\

$c_3$       & $-3.4\,\mbox{GeV}^{-1}$  & See \cite{qcdsf-mn} &
$F_0$       & $0.0862\mbox{GeV}$       & \cite{Colangelo:2003hf} &
$\langle\Delta x\rangle^{(u-d)}$ & $0.21$ & \cite{Dorati:2007bk} \\

$c_A$       & $1.5$                    & \cite{Gail:2005gz} &
$M_0$       & $0.889\mbox{GeV}$        & \cite{Dorati:2007bk} &
& & \\

\hline
\et
\ec
\caption{\label{tab:phen}%
Phenomenological values used as input to our fits.}
\end{table}

To extrapolate our results for the mass of the delta $\mdelta(\mps)$
we fit these to the second order small scale expansion (SSE) expression
\cite{Bernard:2005fy}
\be
m_\Delta(m_{\rm PS}) = M_{\Delta,0} - 4 a_1 \mps^2 -
\frac{3}{32\pi F_0^2} \frac{25 h_A^2}{81} \mps^3,
\label{eq:mdelta}
\ee
where we fixed $F_0$
to the value given in Table~\ref{tab:phen}.  We find that
$\mdelta(\mps=m_{\pi})$ is equal to the experimental value within
statistical errors. The other fit parameters are consistent with
expectations \cite{Bernard:2005fy}: $a_1 = -0.8(3)\,\mbox{GeV}^{-1}
\simeq c_1$ and $h_A = 1.5(4) \lesssim 9 g_A/5$.

\section{Nucleon axial coupling}

We will now consider the form factor of the nucleon axial current $G_A(Q^2)$
at zero momentum.\footnote{Results for non-zero momentum have been
presented by W. Schroers at this conference \cite{qcdsf-ff}.} The axial
coupling constant $g_A = G_A(0)$ is determined from the renormalised
axial vector current $A_{\mu}^R = Z_A\,(1+b_A\,a m_q) A_{\mu}$, where
$a m_q = (1/\kappa - 1/\ksc)/2$. Here we only consider the iso-vector case
where contributions from disconnected terms cancel.  While $Z_A$ is known
non-perturbatively \cite{qcdsf-ga}, $b_A$ is only known perturbatively
and is computed using tadpole improved one-loop perturbation theory.
Note that for forward matrix elements there is no need for improvement
of $A_{\mu}$.

The quark mass dependence of the iso-vector nucleon axial coupling
$g_A(\mps)$ has been calculated using Heavy Baryon $\chi$PT (HB$\chi$PT)
\cite{qcdsf-ga}.  These calculations have been performed both in the
infinite volume limit as well as for a finite spatial cubic box of
length $L$.
We define
$g_A(\mps, L) = g_A(\mps) + \Delta g_A(\mps, L)$,
where $\Delta g_A(\mps, L)$ denotes the finite size effects.

\begin{figure}[t]
\bc
\includegraphics[scale=0.55]{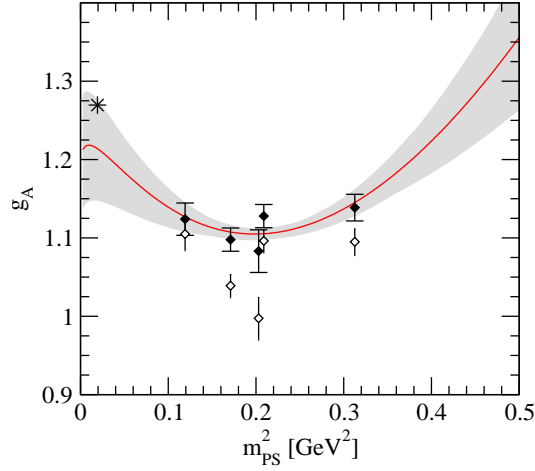}
\vspace*{-5mm}
\ec
\caption{\label{fig:gA}%
Lattice results for $g_A$ together with a fit to an expression
obtained from HB$\chi$PT. The filled symbols show the lattice
results after finite size effects have been corrected.}
\end{figure}

We fit our data to the results obtained
in~\cite{qcdsf-ga} fixing the values for $F_0$, the leading
axial $N\Delta$ coupling $c_A$ and the
$N\Delta$ mass splitting at the physical point $\Delta_0$
to the phenomenological
values given in Table~\ref{tab:phen}. We furthermore set
$B_{20}^r(\lambda)^{SSE}=0$ and use $\lambda = m_\pi$.
A comparison of our data and the resulting fit is shown in
Fig.~\ref{fig:gA}.

The raw lattice data is significantly smaller than the experimental
value. The fit to the HB$\chi$PT expression
suggests that this might be due to significant finite size
effects and a rather strong quark mass dependence in the range
$0 < \mps \lesssim 300\,\mbox{MeV}$.
$g_A(\mps=m_\pi)$ is found to be smaller than the experimental
value, which is however not significant given the
large statistical errors.
The other fit parameters are of natural size.

\section{Moments of unpolarised structure functions}

Finally, we consider the lowest moment of the unpolarised nucleon structure
functions, $\langle x\rangle = A^q_{2,0}(0)$, where $A^q_{2,0}$ is the
first moment of the PDF $H^q(x,\xi,Q^2)$ at $\xi=0$.%
\footnote{Results for $A^q_{2,0}$ at $Q^2 > 0$ have been presented by
M.~Ohtani at this conference \cite{munehisa}.} This moment is determined
from the matrix element
\be
\langle N(\vec{p}) |
\left[
\overline{u}\,\gamma^{\{\mu_1} \Dlr^{\mu_2\}}\,u
\right] | N(\vec{p}) \rangle =
2 A^q_{2,0}
\left[p^{\mu_1} p^{\mu_2}\right].
\ee
For $O(a)$-improvement of the operator we use the perturbative
results obtained in \cite{Capitani:2000xi}, where we inserted the
boosted coupling constant $(g^*)^2 = g^2 / u_0^4$.
The unknown improvement coefficients are set to zero as the corresponding
improvement operators turn out to be small.
The renormalisation is both scale and scheme dependent. We have calculated
the renormalisation constant $\Delta Z^{\rm lat}_{\rm v2b}(a)$, which
translates our lattice results into RGI, non-perturbatively using the
RI'-MOM method. For comparison with other results we have to convert
our numbers to $\msbar$.
The corresponding factor
$\left(\Delta Z^{\msbar}_{\cal O}(\mu=2{\rm GeV})\right)^{-1}$
is calculated perturbatively using
$\Lambda^{\msbar} = 261(17)(26)\,\mbox{MeV}$ 
\cite{Gockeler:2005rv}.

The lattice results for $\langle x\rangle$ in the iso-vector case have
been found to be significantly larger than the experimental value.
It had been suggested that this quantity may become
much smaller at very light quark masses \cite{Detmold:2001jb}.
This has been confirmed by recent calculations in the framework of
baryon chiral perturbation theory (B$\chi$PT)
\cite{Dorati:2007bk}.

\begin{figure}[t]
\bc
\includegraphics[scale=0.5]{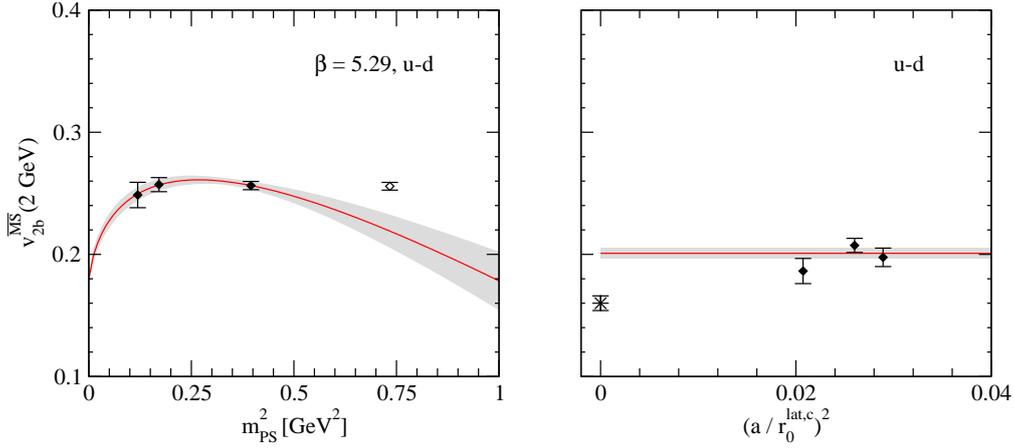}
\vspace*{-5mm}
\ec
\caption{\label{fig:x-umd}%
The left plot shows our results for $\langle x\rangle^{(u-d)}(\mps)$ at
$\beta=5.29$ (the results for other values of $\beta$ are similar). The
solid line denotes
a fit to an ansatz obtained from B$\chi$PT \cite{Dorati:2007bk}.
The right plot shows $\langle x\rangle^{(u-d)}(\mps=\mpi)$ as
a function of the lattice spacing together with a fit to a constant.}
\vspace*{-2mm}
\end{figure}

\begin{figure}[t]
\bc
\includegraphics[scale=0.5]{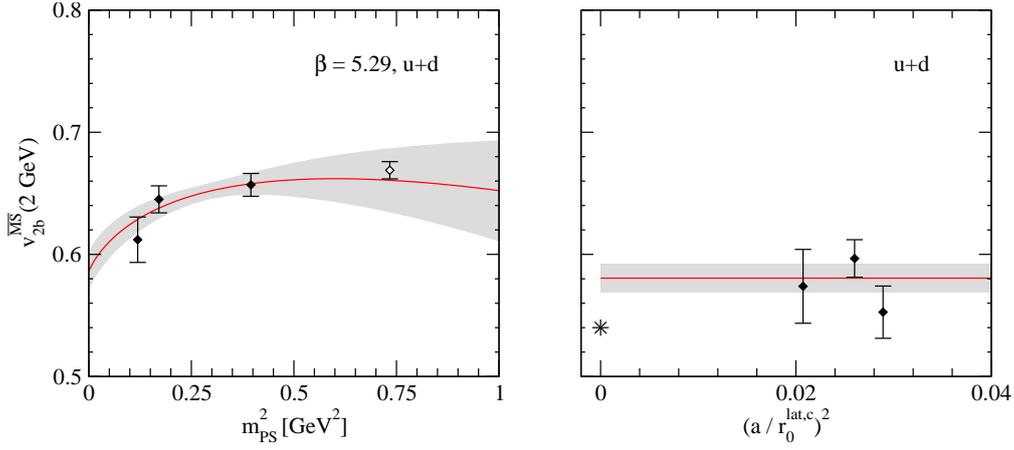}
\vspace*{-5mm}
\ec
\caption{\label{fig:x-upd}%
Same as Fig.~\protect\ref{fig:x-umd} but for the iso-scalar case
$\langle x\rangle^{(u+d)}$.}
\vspace*{-1mm}
\end{figure}

We compare our lattice results with these calculations both in the
iso-vector case ($u-d$) and the iso-scalar case ($u+d$). Note that the
latter might be affected by contributions from disconnected terms
which we have not calculated so far. In our fits we use phenomenological
input for the pion decay constant $F_0$, the nucleon mass $M_0$ and
the nucleon axial coupling $g_{A,0}$ as listed in
Table~\ref{tab:phen}. As suggested in \cite{Dorati:2007bk} we assume
the coupling $\Delta a^{(u-d)}_{2,0} \simeq \langle\Delta x\rangle^{(u-d)}$.
We finally end up with two free fit parameters, i.e.
$a^{u-d}_{2,0}$ and
$c_8(\lambda = 1\,\mbox{GeV})$ as well as
($a^{u+d}_{2,0}$) and
$c_9(\lambda = 1\,\mbox{GeV})$
in the iso-vector and iso-scalar case, respectively.
With the statistical errors and the number of fit parameters
being sufficiently small we fit the data for different values of $\beta$
separately restricting the fit range to $0 < \mps \lesssim 650\,\mbox{MeV}$
(see Fig.~\ref{fig:x-umd} and \ref{fig:x-upd}).

In the iso-vector case we still find little indication for
$\langle x\rangle^{(u-d)}(\mps)$ becoming smaller for
$\mps \rightarrow \mpi$. Currently this is not in contradiction to
the B$\chi$PT results since quark masses may
still be too large. Furthermore, results at light quark masses may be
affected by finite size effects. Within statistical errors the 
discretisation effects seem to be small. In the iso-scalar case we
find evidence for $\langle x\rangle^{(u+d)}(\mps)$ to become smaller
at lighter quark masses. Our results for $\mps=\mpi$ are close to
the phenomenological value.

\section{Conclusions}

In this paper we presented recent progress
in improving our control on the systematic
errors for various quantities in the baryon sector. Finite size effects
were found to be small in the case of the nucleon mass $\mN$ and large
in the case of the nucleon axial coupling $g_A$. Investigation of such effects
for the delta mass $\mdelta$ and the lowest moment of the unpolarised
nucleon structure functions $\langle x\rangle$ is still pending.
For all quantities considered here discretisation effects seem to be small,
although it remains difficult to distinguish potential $O(a^2)$ effects from
other systematic errors.
The largest uncertainties stem from the extrapolation of our results to
the physical quark masses. We utilised results obtained in $\chi$PT
to perform these extrapolations. Although most of our lattice simulations
have been performed at quark masses outside the region where
$\chi$PT is expected to converge, our fit parameters turned out to be
of natural size and consistent with what is expected from phenomenology.

To get better control on the quark mass dependence, simulations at
significantly lighter quark mass are mandatory. Fortunately, with recent
algorithmic improvements and even faster computers becoming available
exploring the quark mass region with $\mps \lesssim 300\,\mbox{MeV}$
starts to become feasible.

\section*{Acknowledgements}
\vspace*{-2mm}

The numerical calculations have been performed on the Hitachi SR8000 at
LRZ (Munich), the Cray T3E at EPCC (Edinburgh) \cite{Allton:2001sk} the
APE{\it 1000} and apeNEXT at NIC/DESY (Zeuthen), the BlueGene/L at NIC/FZJ
(J\"ulich) and EPCC (Edinburgh). Some of the configurations at the small
pion mass have been generated on the BlueGene/L at KEK by the Kanazawa
group as part of the DIK research programme.  This work was supported in
part by the DFG, by the EU Integrated Infrastructure Initiative Hadron
Physics (I3HP) under contract number RII3-CT-2004-506078.


\vspace*{-2mm}


\begin{thebibliography}{99}

\bibitem{gerrit}
  G.~Schierholz {\it et al.} [QCDSF Collaboration],
  \pos{PoS(LATTICE 2007)133}.

\bibitem{qcdsf-mn}
  A.~Ali Khan {\it et al.}  [QCDSF-UKQCD Collaboration],
  Nucl.\ Phys.\  B {\bf 689} (2004) 175.

\bibitem{qcdsf-ga}
  A.~Ali Khan {\it et al.} [QCDSF Collaboration],
  Phys.\ Rev.\  D {\bf 74} (2006) 094508.

\bibitem{qcdsf-ff}
  M.~G\"{o}ckeler {\it et al.} [QCDSF Collaboration],
  Phys.\ Rev.\ D {\bf 71} (2005) 034508;
  W.~Schoers {\it et al.} [QCDSF Collaboration],
  \pos{PoS(LATTICE 2007)161}.

\bibitem{Gail:2005gz}
  T.~A.~Gail and T.~R.~Hemmert,
  Eur.\ Phys.\ J.\  A {\bf 28} (2006) 91.

\bibitem{Eidelman:2004wy}
  S.~Eidelman {\it et al.}  [Particle Data Group],
  Phys.\ Lett.\  B {\bf 592} (2004) 1.

\bibitem{Colangelo:2003hf}
  G.~Colangelo and S.~D\"{u}rr,
  Eur.\ Phys.\ J.\  C {\bf 33} (2004) 543.

\bibitem{Dorati:2007bk}
  M.~Dorati, T.~A.~Gail and T.~R.~Hemmert,
  \texttt{arXiv:nucl-th/0703073}.

\bibitem{Hemmert:2003cb}
  T.~R.~Hemmert, M.~Procura and W.~Weise,
  Phys.\ Rev.\  D {\bf 68} (2003) 075009.

\bibitem{Bernard:2005fy}
  V.~Bernard, T.~R.~Hemmert and U.~G.~Meissner,
  Phys.\ Lett.\  B {\bf 622} (2005) 141.

\bibitem{Bernard:2007zu}
  V.~Bernard,
  \texttt{arXiv:0706.0312} (\texttt{hep-ph}).

\bibitem{munehisa}
  M.~Ohtani {\it et al.} [QCDSF Collaboration],
  \pos{PoS(LATTICE 2007)158}.

\bibitem{Gockeler:2005rv}
  M.~G\"{o}ckeler {\it et al.} [QCDSF Collaboration],
  Phys.\ Rev.\  D {\bf 73} (2006) 014513.

\bibitem{Capitani:2000xi}
  S.~Capitani {\it et al.} [QCDSF Collaboration],
  Nucl.\ Phys.\  B {\bf 593} (2001) 183.

\bibitem{Detmold:2001jb}
  W.~Detmold {\it et al.},
  Phys.\ Rev.\ Lett.\  {\bf 87} (2001) 172001.

\bibitem{Allton:2001sk}
  C.~R.~Allton {\it et al.}, 
  Phys.\ Rev.\ D {\bf 65} (2002) 054502.

\end{thebibliography}
\end{document}